\documentclass[]{aa}

\usepackage{graphicx,color,txfonts,longtable,rotating,times} 
\usepackage[]{natbib}

\newfont{\gwpfont}{cmssq8 scaled 1000}
\newcommand{\rtwo}{{\gwpfont REFLEX~II}}
\newcommand{\smass}{$h^{-1}$M$_{\odot}$}
\newcommand{\sthree}{$10^{13}h^{-1}$M$_{\odot}$}
\newcommand{\sfour}{$10^{14}h^{-1}$M$_{\odot}$}
\newcommand{\cs}{\textit{CS}}
\newcommand{\gcs}{\textit{GCS}}

\begin{document}

\title{Characterising superclusters with the galaxy cluster distribution}
\author{Gayoung Chon\inst{1}, Hans B\"ohringer\inst{1}, Chris A. Collins\inst{2},
Martin Krause\inst{3,1}}
\offprints{Gayoung Chon, gchon@mpe.mpg.de} 
\institute{$^1$ Max-Planck-Institut f\"ur extraterrestrische Physik, 
85748 Garching, Germany\\
$^{2}$Astrophysics Research Institute, 
Liverpool John Moores University, IC2, Liverpool Science Park, 
146 Brownlow Hill, Liverpool L3 5RF, UK\\
$^3$ Excellence Cluster Universe, Boltzmannstrasse 2, D-85748 Garching, Germany
}
\date{Submitted April 2014; Accepted June 2014}

\abstract
{
Superclusters are the largest, observed matter density structures in the Universe. 
Recently \cite{chon13} presented the first supercluster catalogue constructed with 
a well-defined selection function based on the X-ray flux-limited cluster survey, \rtwo. 
For the construction of the sample we proposed a concept to find the large objects with a minimum 
overdensity such that most of their mass will collapse in the future. 

The main goal of the paper is to provide support for our concept using simulations that we can, 
on the basis of our observational sample of X-ray clusters, construct a supercluster sample defined 
by a certain minimum overdensity, and to test how superclusters trace the underlying dark matter 
distribution. 
Our results confirm that an overdensity in the number of clusters is tightly correlated with 
an overdensity of the dark matter distribution. 
This enables us to define superclusters such that most of the mass will collapse in the future 
and to get first-order mass estimates of superclusters on the basis of the properties of 
the member clusters. 
We also show that in this context the ratio of the cluster number density and dark matter mass density 
is consistent with the theoretically expected cluster bias.  

Our previous work provided evidence that superclusters are a special environment for 
density structures of the dark matter to grow differently from the field as characterised 
by the X-ray luminosity function. Here we confirm for the first time that this originates from 
a top-heavy mass function at high statistical significance provided by a Kolmogorov-Smirnov test. 
We also find in close agreement with observations that the superclusters occupy only 
a small volume of few percent while they contain more than half of the clusters in the present 
day Universe. 
}


\authorrunning{Chon et al.} 
\titlerunning{Characterising superclusters with the galaxy cluster distribution}

\maketitle

\section{Introduction}

Superclusters are the largest, prominent density enhancements in our Universe. 
They are generally defined as groups of two or more galaxy clusters above a given spatial density 
enhancement~\citep{bahcall-88}. 
Their sizes vary between several tens of Mpc up to about $150$~$h^{-1}$~Mpc. 
As the time a cluster needs to cross a supercluster is larger than the age of the Universe, superclusters 
cannot be regarded as relaxed systems. 
Their appearance is irregular, often flattened, elongated or filamentary and generally not spherically symmetric. 
This is a sign that they still reflect, to a large extent, the initial conditions set for the structure formation 
in the early Universe. 
They do not necessarily have a central concentration, and are without sharply defined boundaries.

The first evidence of superclusters as agglomerations of rich clusters of galaxies was provided by \citet{abell-61}. 
The existence of superclusters was confirmed by \citet{bogart-73}, \citet{hauser-73} and \citet{peebles-74}. 
So far numerous supercluster catalogues have been published based essentially on optically selected samples of 
galaxy clusters. 
Several supercluster catalogues based on samples of Abell/ACO clusters of galaxies followed, e.g. \citet{rood-76}, 
\citet{thuan-80}, \citet*{bahcall-84}, \citet*{batuski-85}, \citet{west-89}, \citet{zucca-93}, \citet*{kalinkov-95}, 
\citet{einasto-94, einasto-97}, and \citet{einasto-01}, among others.
\citet{einasto-01} was the first to use X-ray selected clusters as well as Abell clusters. 
X-ray selected galaxy clusters are very good tracers of the large-scale structure as their X-ray luminosity 
correlates well with their mass unlike optically selected samples of galaxy clusters, e.g.~\cite{pratt09}. 
Such samples suffer, among other things from projection effects. 
However, none of these catalogues is based on cluster catalogues with a well-understood selection criterion. 
This means that they can neither be used as a representative sample to study the properties of superclusters 
nor easily be compared to simulations. 

\cite{chon13} made a fresh approach in this respect by presenting the first statistically well-defined supercluster 
catalogue based on the \rtwo\ (extended ROSAT-ESO Flux Limited X-ray) cluster survey~\citep{chon12,r2const} using 
a friends-of-friends algorithm to construct the superclusters. 
The \rtwo\ cluster catalogue comes with a well-understood selection function and it is complete, homogeneous and 
of very high purity. 
This allows a relatively straightforward reconstruction of the survey with simulated data.
\defcitealias{chon13}{C13}\citet{chon13}, hereafter \citetalias{chon13}, presented the construction of superclusters 
and studied observed properties of the X-ray superclusters. 
The strategy of the supercluster sample construction in this previous paper was based on the assumption that 
choosing a certain linking length for the cluster association to superclusters we obtain superclusters which 
have a minimum dark matter overdensity. 
By choosing the right linking length value in relation to the mean cluster density we conjectured that we can 
obtain a supercluster sample including those objects that marginally collapse in the future. 
This hypothesis relies on two assumptions, 
(i) a certain linking length corresponds to a certain number overdensity of clusters, and (ii) an overdensity 
in the number of clusters above a certain mass limit corresponds to a closely correlated value of the overdensity 
in the dark matter for the supercluster volume. 

The main aim of the present paper is to use simulations to test these crucial assumptions, and to examine 
how superclusters trace the underlying dark matter distribution.  
Using the dark matter distribution of the Millennium simulation~\citep{springel05} we compare the cluster density 
with the dark matter density in superclusters, 
constructed with a friends-of-friends algorithm such that they should have a minimum spherical overdensity.
To find out the way superclusters trace the dark matter we consider two quantities. 
We first study the correlation between the mass fraction of a supercluster represented in its member clusters 
to the total mass of a supercluster probed by all halos in the Millennium Simulation, and investigate how the 
cluster number density is correlated with the matter overdensity of a supercluster, analogous to the bias of 
the power spectrum of clusters, and examine how simulated superclusters trace the underlying matter distribution. 

According to Birkhoff's theorem structure evolution in a supercluster region can be modelled in an equivalent 
way to a Universe with a higher mean density than that of our Universe. 
The major difference between these two environments is then a slower growth of structure in the field compared 
to the denser regions of superclusters in the recent past. 
Thus, we would expect a more top-heavy X-ray luminosity function in superclusters. 
Among the properties of the superclusters investigated in~\citetalias{chon13} two major findings were:  
(1) that the volume occupied by those superclusters is very small, two per cent, while more than half of the 
\rtwo\ clusters are found in the superclusters and (2) that the cumulative X-ray luminosity function of 
the Volume-Limited-Sample (VLS) of the \rtwo\ superclusters differs from that in the field supported by 
a Kolmogorov-Smirnov (KS) test. 
Since the X-ray luminosity of clusters is tightly correlated with their mass, we may expect that there are 
more massive clusters in superclusters than in the field. 
In~\citetalias{chon13} we also discussed an alternative explanation for the top-heavy luminosity function, 
that the luminosity of clusters is temporarily boosted in a dense region due to an increased rate of merger events. 
In this case we would not find correspondingly more massive clusters in superclusters. 
Our study with the Millennium Simulation will allow us to distinguish between these hypotheses in this paper.
There are previous studies which investigated various quantities within the supercluster environment, 
e.g.~\citet{einasto03_ref,einasto05_ref,einasto12_ref,luparello13_ref}, with the galaxy survey data 
or N-body simulations, and the future evolution of superclusters were studied 
by~\citet{duenner06,araya09,luparello11_ref}. 
They found that in the regions of superclusters there are more richer and massive systems, 
enhanced velocity dispersions, and larger star formation rates, which all point
to the suggestion that the superclusters provide a different environment for structures to grow
in comparison to the field. 
Our approach is different from the previous work in the way that we select the clusters
of galaxies as tracer objects to probe superclusters, and that they are selected closely
following the flux-limited X-ray observation. 

This paper is organised as follows. Sec. 2 describes the construction of superclusters from the Millennium 
simulation, and their basic properties. 
The connection of the linking length to the cluster density ratio is explored in Sec. 3. 
We investigate how superclusters trace the dark matter distribution by studying the mass fraction of clusters and 
the bias introduced by the cluster number overdensity in Sec. 4. 
We probe the supercluster environment by comparing the cluster mass function and volume fraction of superclusters 
to those of the field in Sec. 5. We conclude in Sec. 6 with some discussions.

\section{Construction of superclusters in simulation}

We use the Millennium Simulation to study the properties of superclusters constructed with 
the Dark Matter (DM) halos. 
It contains a total of 2160$^3$ particles with a mass of 8.6$\times10^{8}$\smass\ 
in a box size of 500 $h^{-1}$Mpc. 
It adopts a flat-$\Lambda$CDM cosmology with $\Omega_m$=0.25, $h$=0.73, $\sigma_8$=0.9. 

Our aim with the simulation is to explore if superclusters can be used to probe the 
statistics of the large-scale structure in our Universe, and to test out the concept 
of constructing superclusters with a certain minimum overdensity of clusters.
To construct the supercluster catalogue with the Millennium simulation data analogous  
to typical X-ray cluster observations we define clusters in the DM simulation as the halos 
above a given mass limit. 
We adopt M$_{200}$ as the halo mass, where M$_{200}$ is defined as the mass inside the radius 
where the mean halo density is 200 times the mean density of the universe. 
We are interested in the mass range above \sthree\, which encompasses the whole range of 
clusters and groups of galaxies. 
For better discussion two reference catalogues are considered with two lower mass limits 
of \sfour\ and \sthree. 
Since current lower mass limits of cluster surveys are typically about \sfour\, we define 
the halos of mass above \sfour\ as clusters, of which our main cluster catalogue is composed. 
The second main catalogue that we consider is made with the halos of mass above \sthree. 
These halos correspond to clusters and groups of galaxies, which we take as a representative 
sample for future cluster and group surveys. 
Hence we will refer to the supercluster catalogue built with the halos of mass above \sfour\ as \cs\ 
and that above \sthree\ as \gcs\ for brevity. 
We restrict our study here to a snapshot at $z=0$. 

With the definition of clusters given above we construct a catalogue of superclusters 
with a friends-of-friends (fof) algorithm as~\citetalias{chon13}. 
The key parameter in this algorithm is the linking length, within which friends are found,  
and this parameter determines the types of superclusters that are constructed. 
We adopt the definition of a linking length in~\cite{zucca-93}, which depends on the overdensity, 
$f$, i.e. the cluster density enhancement over the mean cluster density, $f=n/n_o$. 
As the local density $n \propto l^{-3}$, the linking length is inversely proportional to 
the density of the sources in the volume, $n_o$, and we define it as $l=(n_of)^{-1/3}$.
Hence the choices of $f$ affect the definition of superclusters. 
In this paper $f$ is fixed to ten for the reasons detailed in~\citetalias{chon13}, 
and we only summarise here briefly. 
We have calculated this required overdensity value by assuming a spherical top-hat collapse 
model for the superclusters and by integrating the Friedmann equations describing the evolution
of the superclusters given 
$\Delta_{\rm{CL,m}}=[(\Delta_{\rm{DM,c}}+1)/\Omega_{\rm{m}} -1] b_{\rm{CL}}$. 
$\Delta_{\rm{CL,m}}$ is the cluster overdensity against the mean density, 
$\Delta_{\rm{DM,c}}$ the DM overdensity against the critical density 
in the superclusters, and $b_{\rm{CL}}$ is the bias in the density fluctuations
of clusters with the condition $\Delta_{\rm{DM,c}} \gtrsim 1.4$.
Thus for superclusters at low redshifts that will collapse in the future 
we find the following condition, $\Delta_{\mathrm{CL},m} \ge $ 17-35 for a cluster bias of 2-4. 
As noted in~\citetalias{chon13} we sample superstructures more comprehensively 
by taking $f$=10, which include those slightly beyond definitely bound structures. 
Therefore this linking parameter is rather conservative by including slightly
more material than what is gravitationally bound in a $\Lambda$CDM Universe.
We can see that most of the superclusters are still recovered with $f$=50~\citepalias{chon13}
that the majority of the matter in the superclusters will collapse in the future, 
and in addition it was found that a volume of $f$=10 selects superclusters that 
correspond well to superclusters described in the literature. 
Our search for friends of a supercluster starts from the brightest clusters 
in the halo catalogue, where we call the brightest cluster in a supercluster BSC.
We note that the definition of superclusters in our work also includes those objects 
with two cluster members.

\subsection{Physical properties}

There are two main supercluster catalogues constructed with the recipes described above. 
The \cs\ contains 3569 clusters from which 607 superclusters are constructed while 
there are 8893 superclusters out of 51528 clusters in the \gcs.
In the following we consider two physical properties that result from the choice of
the overdensity parameter, $f$.

We note that~\cite{einasto-sim} compared the properties of superclusters from the 2dF 
galaxy redshift survey~\citep{einasto-07} to those constructed with the Millennium Simulation. 
Their main conclusion was that most of the properties of the 2dF superclusters 
are in good agreement with those of the Millennium superclusters except for the 
luminosity and multiplicity distributions. 
Unfortunately a direct comparison to their results is not possible due to the fact that 
our supercluster catalogue is constructed from clusters directly, and their catalogue on 
galaxy overdensities.

\subsubsection{Multiplicity function}

The multiplicity function is defined as the number of member clusters in a supercluster, equivalent to the richness 
parameter of an optical cluster of galaxies. 
We define the superclusters as systems with two or more member clusters in our study, and 
systems with two member clusters are referred to as pair superclusters. 

\begin{figure}
  \resizebox{\hsize}{!}{\includegraphics{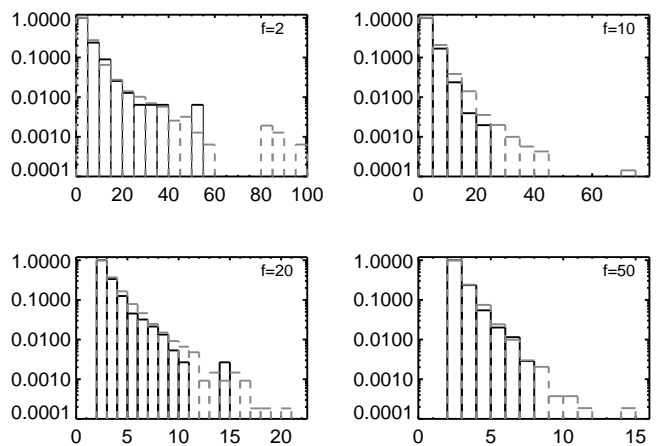}}
  \caption{Normalised multiplicity function of the superclusters 
    for a range of overdensity parameters, $f$, constructed
    with the halos in the \cs\ (black solid line), and with those
    in the \gcs\ (gray dashed line).
    The value of $f$ is written in each panel. 
    Richer systems are abundant as $f$ gets smaller, 
    which is equivalent to a larger linking length 
    (upper left panel), while less rich systems are more 
    common with larger $f$ values (lower right panel). 
  }
  \label{fig:mult}
\end{figure}

The normalised multiplicity function of the superclusters is shown in Fig.~\ref{fig:mult} for a range of 
overdensity parameters, $f$. 
Those constructed with the \cs\ are shown in black solid line, and those with the \gcs\ in gray dashed line. 
We note that there are three very rich superclusters with $\textgreater$ 100 members or more in the $f$=2 case
in both \cs\ and \gcs, which are not shown. 
For a fixed $n_o$ a larger value of $f$ corresponds to a smaller linking length, and in this case 
there are fewer very rich superclusters, and less rich systems dominate the distribution. 
On the other hand with smaller values of $f$ the number of very rich superclusters increases due to a large
linking length. 
In either case the histograms of the multiplicity for all $f$ are dominated by pair and less rich superclusters 
and the spread of the multiplicity becomes larger with a larger linking length. 
We see these trends in Fig.~\ref{fig:mult}, and there is a good agreement in the shape of the multiplicity 
functions of supercluster samples constructed with different mass limits.
The Kolmogorov-Smirnov (KS) test in each case of Fig.~\ref{fig:mult} gives the probability of unity meaning 
that it is highly likely that the multiplicity distributions formed by both \cs\ and \gcs\ originate from 
the same parent distribution.

\subsubsection{Extent}

Superclusters are the largest objects that are seen in, for example, galaxy redshift surveys, 
and we are interested to find how large these prominent features of the large-scale structure are. 
We define the extent of a supercluster by the maximum distance between 
the centre of mass of a supercluster to its furthest member clusters. 

\begin{figure}
  \resizebox{\hsize}{!}{\includegraphics{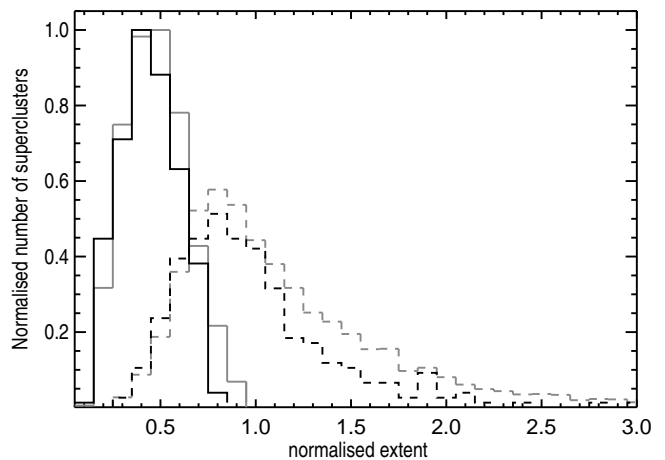}}
  \caption{
    Distribution of the normalised extent of the superclusters.
    The superclusters in the \cs\ is shown in black, and those
    in the \gcs\ in gray.
    Superclusters with three member clusters or more are marked 
    by a dashed line, and pair superclusters as solid line. 
  }
  \label{fig:extent}
\end{figure}

Fig.~\ref{fig:extent} shows the distribution of the supercluster extent for the two nominal mass limits. 
The extent is normalised by the linking length of the corresponding catalogue, which is 15.2~Mpc/h for 
the \cs\, and 6.2~Mpc/h for the \gcs. 
The pair superclusters are denoted by solid lines, and the rest of the richer systems in dashed lines.  
Less than a per cent of superclusters in the \gcs\ are left out in the plot with the normalised extent 
larger than three. 
There is a striking similarity between the two samples of the richer superclusters.
The typical size of the richer superclusters is around 80 per cent of the linking length. 
Richer systems show a larger tail towards large extents.
Also in the pair superclusters the distributions are very similar, which is less 
surprising, since in both samples the typical size of pairs is half of the
linking length. This implies that most of the pairs have close to maximum size
for systems of similar mass. 
The similarity between the size distributions of the two samples provides a signature
of the self-similarity of the large-scale structure at these mass and length scales.

\subsubsection{Percolation}

Percolation occurs when structures start being linked together and permeate through the volume.
In our frame of work it can be understood as those systems identified as separate systems 
with a larger overdensity parameter start to connect and form a single super-structure 
as we decrease the overdensity parameter. 
These structures will not collapse to a virialised object in the future.
Hence it is important to understand where the percolation occurs in the sample of superclusters, 
the level of percolation, and how to treat the percolated systems in the analysis. 

\cite{shandarin04_ref} studied the percolation to understand the morphology of 
the supercluster-void network constructed by smoothed density fields in the N-body
simulations. 
They found that the percolating objects make up a considerable fraction of superclusters, 
and that they should be studied separately. 
\cite{liiv-12} also studied the percolation in the supercluster catalogue constructed
with galaxy data depending on the density threshold, which is in some way equivalent
to our overdensity parameter. 

Our strategy in this paper is to build a sample of superclusters with a clear 
selection function of clusters, which resembles the X-ray selection, that will form 
collapsed objects in the future. 
$f$=10 is chosen as a physically motivated overdensity parameter based on solving
the Friedmann equations, and we explained that our nominal catalogues built with $f$=10 
also include those structures that will also partially collapse in Sec. 2. 
Hence it is interesting to investigate how many percolating superclusters there are in 
our sample, and their effect on further analysis in the paper.

Based on the distribution of the multiplicity function we determine the percolation scale, 
which isolates very large structures. For instance the top left panel of Fig.~\ref{fig:mult} 
shows isolated objects beyond multiplicity of 80, and we identify them as percolating objects. 
Hence we define the scale of percolation by the largest system in the multiplicity distribution 
still belonging to the continuous distribution function before the first gap. 
This is slightly more conservative than taking the scale at which the isolation starts, but this 
does not modify our conclusion. 
This categorisation reveals that there are 11 superclusters in the nominal \gcs\, and 3 in the \cs\
that may be the results of percolation, which are very small number fraction of the sample. 
This is reassuring for our choice of the overdensity parameter.

\begin{figure}
  \resizebox{\hsize}{!}{\includegraphics{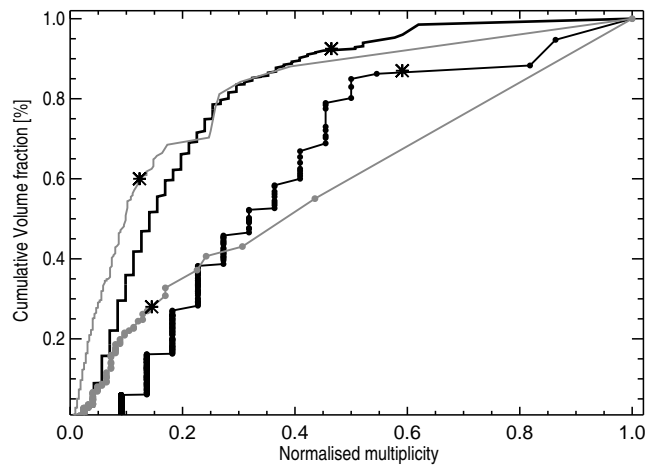}}
  \caption{
    Cumulative volume fraction of superclusters as a function
    of normalised multiplicity. 
    The \cs\ is shown in filled circles, \gcs\ in continuous lines. 
    For the same mass limit the catalogues built with $f$=2 are shown in gray, 
    and $f$=10 in black. 
    Four asterisks mark the estimated starting scale of percolation.
  }
  \label{fig:perc}
\end{figure}

The most evident quantity that percolation affects is the volume fraction of percolating
structures to the total volume of superclusters. 
We show the effect of percolation on the cumulative volume fraction in Fig.~\ref{fig:perc}
as a function of normalised multiplicity. 
Four cases are considered here with the two overdensity parameters, $f$=2 (gray) 
and $f$=10 (black) for the catalogues built with the two mass limits, \sthree\ (line), and 
\sfour (filled circles). 
The percolation scales are marked by asterisks on each curve. 
For the same mass limit percolating objects identified in the $f$=2 catalogue take up much 
larger volume than those in the $f$=10 catalogue. 
In the \sfour\ catalogues they occupy about 70~\% of the total volume for $f$=2
while the fraction reduces to about 15~\% in the nominal $f$=10 catalogue.
Also the percolation starts at smaller normalised multiplicities for a higher mass 
limit as shown by the asterisks. 

We find that with the superclusters built with $f$=10 the contribution of potential percolation 
to the total volume fraction ranges from three to maximum 20~\% for the mass limits between
\sthree\ and \sfour. 
Hence the percolation effect is much less pronounced in our supercluster catalogues in contrast to 
previous findings, e.g.~\cite{shandarin04_ref}. 
We attribute our finding to the fact that clusters of galaxies do not trace long, thin 
Dark Matter filaments as closely as galaxies or the smoothed density field. 
With our nominal catalogues built with $f=$10, we find very few percolating structures, 
hence their contribution to the quantities that are probed in our paper is very limited. 
An extreme treatment of these superclusters is to remove them from the analysis, and as will 
be discussed in relevant sections no major differences are found, except for the total 
volume of superclusters in our analysis.

\section{Assessing supercluster density ratio}

The nature of a supercluster catalogue depends critically on the choice of a linking length in the fof 
algorithm since it determines the types of superclusters that are constructed. 
A too large linking length merges structures that may not be bound gravitationally, on the other hand 
a much smaller value may select only the peaks of larger underlying structures. 
In~\citetalias{chon13} it was shown that the overdensity parameter, $f$, which is inversely proportional to the 
linking length, can be formulated such that the corresponding linking length selects those superclusters that will 
eventually collapse in the future in our standard $\Lambda$CDM universe. 
As we discussed in Sec. 2 we defined a physically motivated linking length, which corresponds to $f$=10 as our 
nominal overdensity parameter to capture systems slightly beyond definitely bound structures based on integrating 
the Friedmann equations.
Our aim in this section is to understand how superclusters are represented by their member clusters constructed 
with this overdensity parameter, based on a spherical top-hat collapse model, via friends-of-friends algorithm. 
In this approach we approximate superclusters to have a spherical shape and for consistency with this theoretically 
motivated characterisation we keep this geometric approximation throughout this paper. 

To answer this question we check the distribution of the actual density ratios of superclusters constructed with 
a given $f$, in retrospect. 
We can measure the density ratio of the superclusters, defined as a ratio of the cluster number density in 
a supercluster over the mean number density of clusters in the volume of consideration. 
This test also provides a sanity check on our assumption, and a census of the typical departure of these 
large structures from the mean density of the universe. 
To calculate the ratio, the volume of a supercluster is assumed to be spherical, and its extent is defined 
in Sec. 2. 

\begin{figure}
  \resizebox{\hsize}{!}{\includegraphics{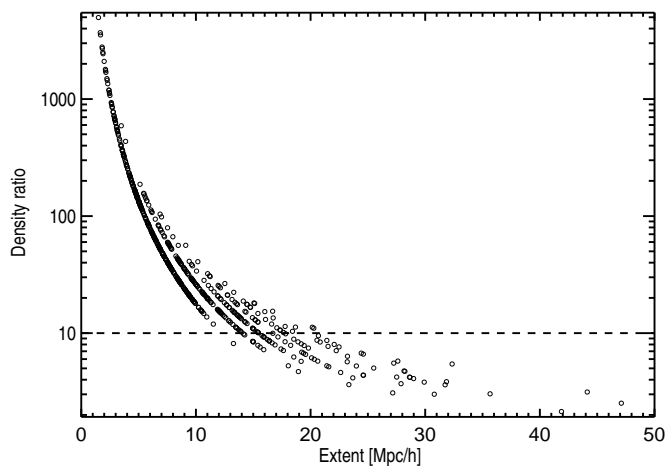}}
  \caption{
    Measured number density ratio of the clusters in the \cs. 
    The cluster number density ratio is defined as a ratio
    between the cluster density in a supercluster and 
    the mean cluster density in the simulation. 
  }
  \label{fig:ov}
\end{figure}

\begin{figure}
  \resizebox{\hsize}{!}{\includegraphics{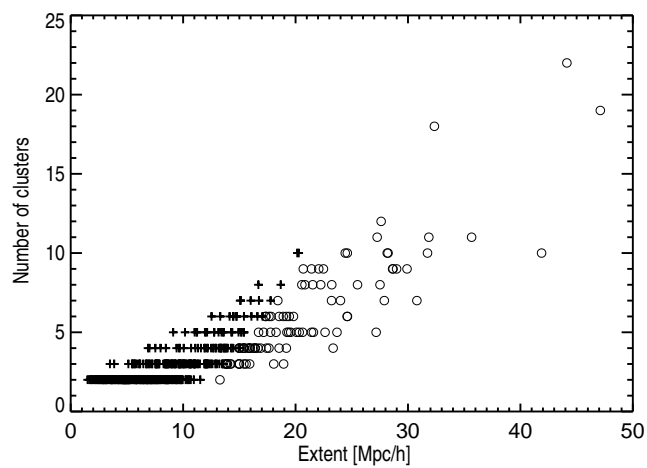}}
  \caption{
    Multiplicity of superclusters shown as a function
    of their extent for the \cs\ as
    in Fig.~\ref{fig:ov}. The open circles represent 
    those superclusters with 
    measured density ratio below ten in Fig.~\ref{fig:ov}
    }
    \label{fig:no-ext}
\end{figure}

\begin{figure}
  \resizebox{\hsize}{!}{\includegraphics{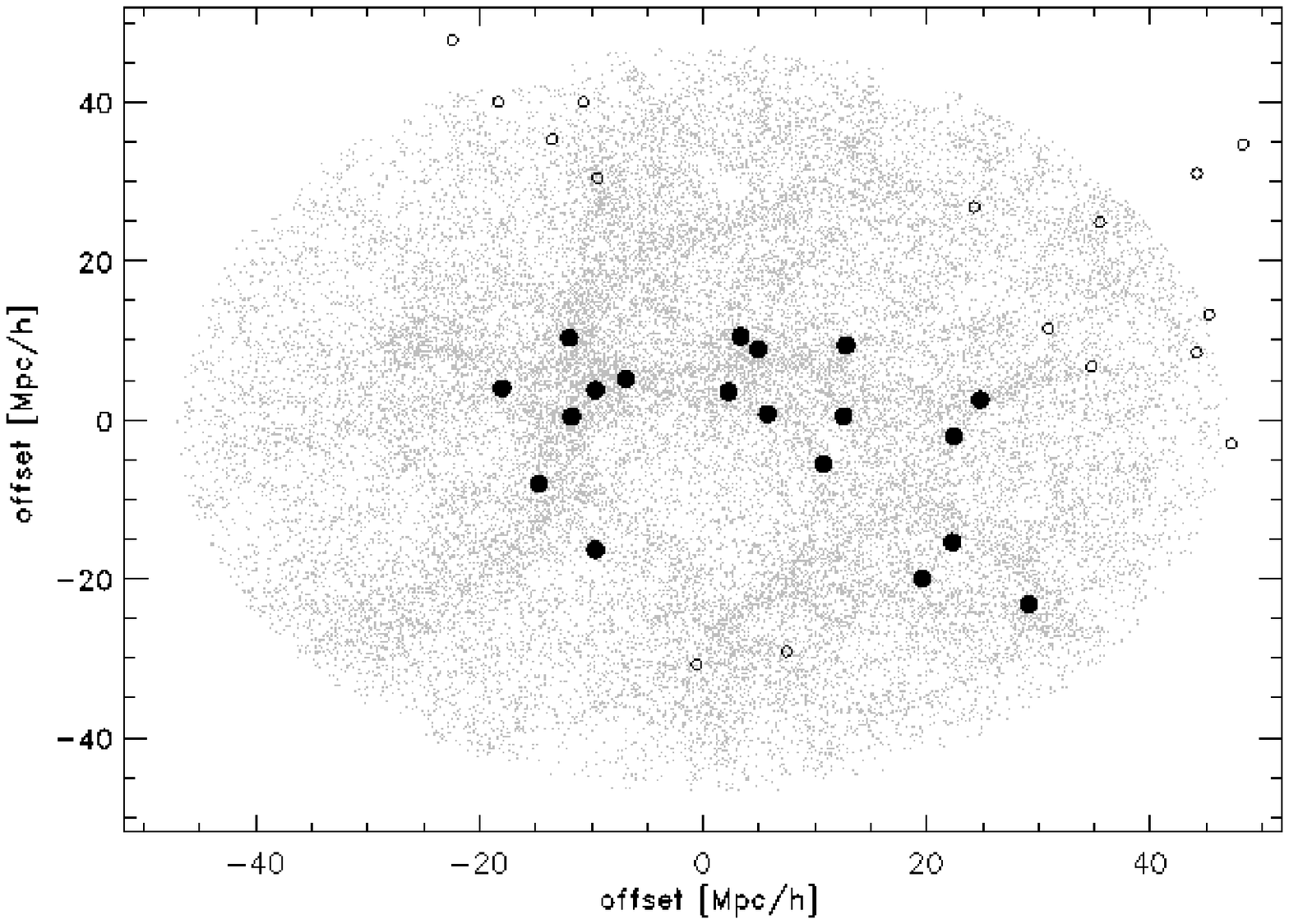}}
  \resizebox{\hsize}{!}{\includegraphics{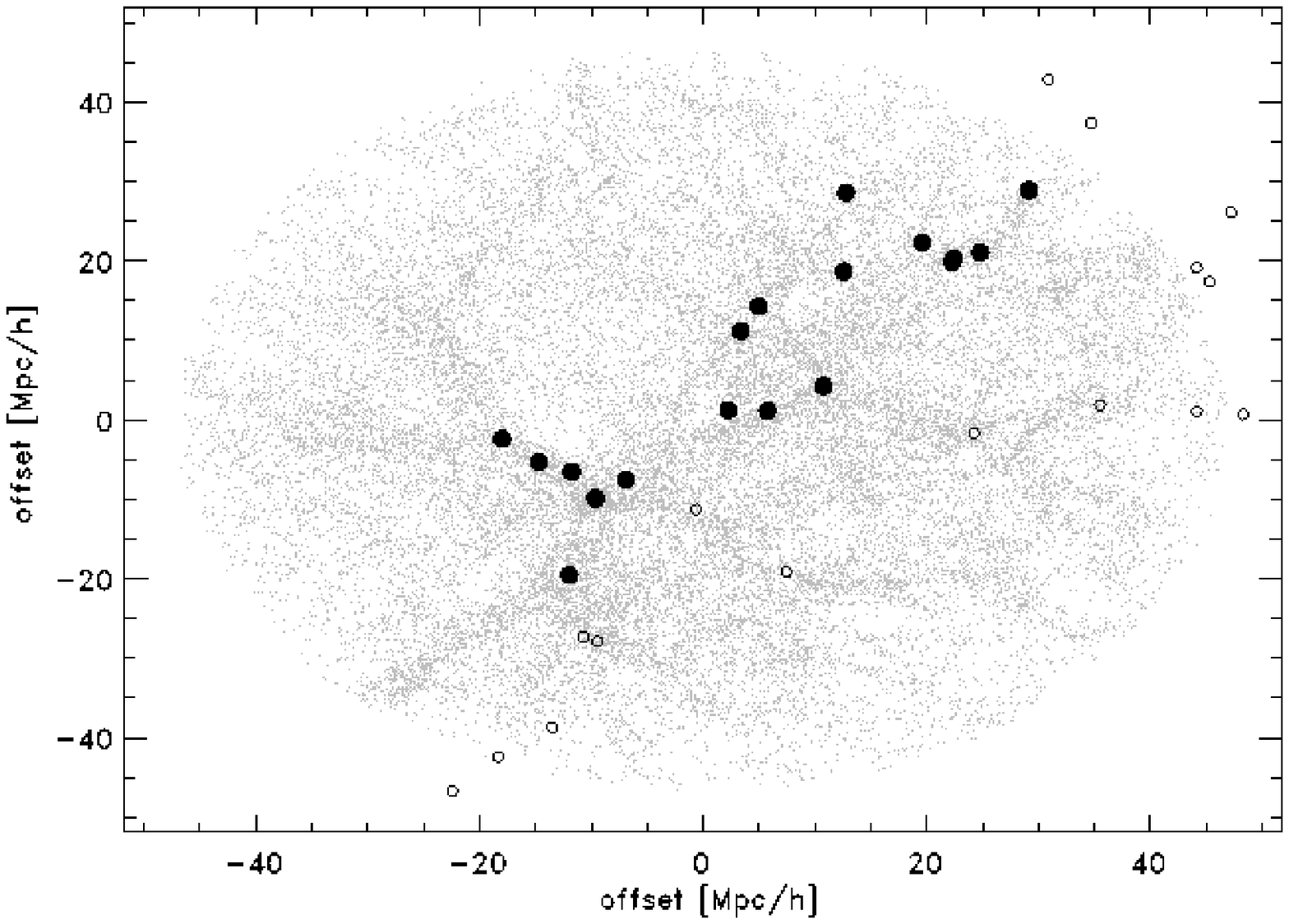}}
  \caption{
      Distribution of all halos in the largest supercluster sliced through 
      two axes, in X-Y (upper), and in X-Z (lower). 
      The offsets are in unit of Mpc/h. The member clusters are shown in filled circles, 
      and all other less massive halos than the mass limit in gray dots where 
      the centre of mass is located at the origin. 
      This supercluster, taken as an example from Fig.~\ref{fig:ov}, represents 
      such superclusters with a measured density ratio less than ten. 
      Typically the member clusters of these superclusters
      form a thin filament along one axis, which causes an over-estimation of a 
      supercluster volume leading to an under-estimation of the measured density ratio.
      The open circles mark the clusters belonging to the neighbouring superclusters 
      whose volumes overlap with the volume of the main supercluster due to the definition of the extent. 
  }
  \label{fig:pan}
\end{figure}

Fig.~\ref{fig:ov} shows the measured density ratio of the \cs\ as a function of the extent. 
The apparent line-like structures are due to the discreteness of the multiplicity function,
since the density ratio is proportional to the multiplicity and the volume of a supercluster.
The dashed line corresponds to the overdensity parameter, $f=10$, with which we constructed the superclusters. 
We note that with a maximum linking length defined by $f$=10, a density ratio of ten is the expected lower limit
with some scatter, as the cluster distances in a particular supercluster can always be smaller than this maximum. 
We see this in particular among the pairs in Fig.~\ref{fig:no-ext} where we find a large number of systems with 
linking lengths very much smaller than the maximum allowed value.
In extreme cases these are systems very close to merging. 
Therefore it is not a surprise to find systems very much above the threshold in Fig.~\ref{fig:ov}.
Also the \cs\ has 9 per cent of superclusters falling below the dashed line allowing 20 per cent scatter 
while for the \gcs\ superclusters, 18 per cent fall below for the same scatter.  
Those superclusters below the overdensity threshold of ten are marked by open circles while the rest of 
superclusters by crosses in Fig.~\ref{fig:no-ext}. 
The former systems are dominantly large in each class of the multiplicity, where the smallest 
system among those has $R$=13.5 Mpc/h in size. 
Fig.~\ref{fig:pan} demonstrates an example of such a system in two projections which has one of the lowest measured 
density ratios. 
Both panels show the member clusters in black filled circles, and all other halos in gray dots 
in the spherical volume defined by the extent of a supercluster. 
The upper panel shows a slice through the X-Y plane, and the lower through the X-Z. 
In this extreme case the clusters are located along a long filament in the X-axis while 
in the other axes the distribution of clusters is rather compact in comparison. 
In this case the spherical volume overestimates the volume that is represented by the member 
clusters. 

Traditionally superclusters have been found with a fof algorithm, and this result reflects 
how well the fof reproduces structures in comparison to a spherical overdensity (SO). 
This is analogous to compiling dark matter halos in N-body simulations which are found by the fof, 
and characterised by the SO method except that superclusters are more complex and are not virialised. 
We find in our analysis that the fof method does a similarly good job in the compilation
of superclusters as for the study of dark matter halos. 

A different point of concern in this context is that superclusters constructed with a fof algorithm 
could potentially have overlapping volume with neighbouring superclusters if their extent is 
defined as in Sec. 2.1.2 with the assumption of a spherical shape. 
Therefore we explored if this occurs in our sample, and show an example in Fig.~\ref{fig:pan}.
Those clusters in the neighbouring superclusters, which have an overlapping volume with this supercluster, 
are shown as open circles. 
For the \cs\ the overlapping volume is 1.5 percent of the total supercluster volume, and for the \gcs\ it is about 
six percent. 
Since the overlapping volume fraction is so small, we neglect it in the further analysis.

\section{Superclusters as Dark Matter tracers}

The fact that the \rtwo\ supercluster sample has been constructed by means of a statistically
well defined sample of closely mass-selected clusters, motivates us to search for a more precise
physical characterisation of the simulated superclusters. We are in particular interested to explore the
relation of the cluster density in superclusters to the underlying dark matter distribution.
This can be studied with DM simulations by applying equivalent criteria to those used
in our X-ray selection. 
In this respect we consider two quantities in this section, the mass fraction of the superclusters represented 
by their member clusters compared to total supercluster mass, and the overdensity of clusters in superclusters 
as a function of the DM mass overdensity. 
A better knowledge of these relations would greatly assist the interpretation of the observations.
Since we have hardly any direct access to the determination of the supercluster masses - neither
dynamical mass estimates nor gravitational lensing studies has yet been applied successfully to 
entire superclusters - an indirect mass estimate would be very helpful. 
Since we have mass estimates of the member clusters through mass-observable scaling relations 
where the X-ray luminosity is the crucial observable in our case, the total supercluster mass can 
be estimated if we can calibrate the cluster mass fraction in superclusters. Similarly we would be 
able to determine the DM overdensity traced by a supercluster if the mass fraction relation or the 
overdensity bias could be calibrated. The following analysis constitutes a first exploration of this territory.

In this section we consider a finer grid of mass limits to form a smoother 
distribution of mass-related quantities between \sthree\ and \sfour where appropriate. 
We note that there are approximately 5-10 per cent of superclusters, depending on the mass 
limit, that will not be included in the samples purely due to the fact that their extent 
in one or more dimensions extends further outside the simulation volume. 
Since this happens the same way for the observations near the survey boundaries, we exclude 
these superclusters. 
Another technical point is the assumption that we make in terms of the DM particle mass 
distribution. We assume that the least massive halos in our study are not biased against the DM. 
This applies also to those particles that do not contribute to the $M_{200}$ of halos.
This is reasonable due to the small mass resolution of particles as well as to the fact that 
unbound particles are spatially distributed equally throughout the simulation. 
In fact the assumption that the halos less massive than the smallest groups have a bias close 
to unity is in the first place based on theory, e.g.~\cite{seljak04}. 
We also tested this by comparing the halo distribution for different halo masses with the dark 
matter density field, and found it to be observed with an accuracy of about 5~\% on a scale 
of 5~Mpc/h or larger.

\subsection{Mass fraction probed by clusters}

From the observations and simulations we have a good understanding of the relation 
between cluster observables and the total mass of a cluster, e.g.~\cite{scalingpap,sheldon09,johnston07}, 
such that the visible components of a cluster represent well the underlying dark matter
component. 
If DM had no preferential scale, this argument should apply in a similar way to superclusters, 
which means that a supercluster mass probed by the total cluster mass should also be correlated 
with the true mass of the supercluster. 
This gave the motivation for this subsection where we investigate if superclusters can be used 
to trace the DM distribution by comparing the total cluster and total DM mass of a supercluster 
in the simulation. 
This is interesting since it leads to the possibility of mass calibrations for 
superclusters just the same way it is done for estimating the total mass of a cluster with 
its observables. 
Our aim here is not to provide an exact fitting formulae to calibrate the masses, but 
to test experimentally how well superclusters are represented in clusters, and to diagnose 
if this is a viable way forward with future cluster survey missions. 
With the definition of the mass fraction as the ratio between the total cluster mass 
and the true supercluster mass, we study what the typical mass fraction of a supercluster is 
that is made up of clusters, and explore if there is any dependency on the mass limit imposed by 
an observation. 
The estimated true mass of a cluster relies on cosmology and a scaling relation in the real 
observation. 
However, using simulations where the true mass is known, we can jump over those 
two constraints directly to calculate the mass fraction. 

As explained above, the true mass of a supercluster is defined to be the sum of all halos 
in the volume with an assumption that the simulation mass resolution is sufficiently small. 
As was mentioned at the beginning of Sec. 3, particles which do not contribute to $M_{200}$ 
are excluded in our work. 
However, we take them into account when calculating the total halo mass of a supercluster as 
a correction factor with the reasonable assumption that these particles are spatially distributed 
in an unbiased way. 
This correction factor, taken as the ratio between the total halo mass and the total particle
mass in the simulation, is 2.02 in our particular case, and is accounted for in the following result. 
For convenience we denote the total cluster mass in a supercluster as CM, and the total mass traced
by halos in the same supercluster as HM such that the mass fraction is defined as 
CM/HM\footnote{Wherever HM is referred to numerically the correction factor is already applied.}.

\begin{figure}
  \resizebox{\hsize}{!}{\includegraphics{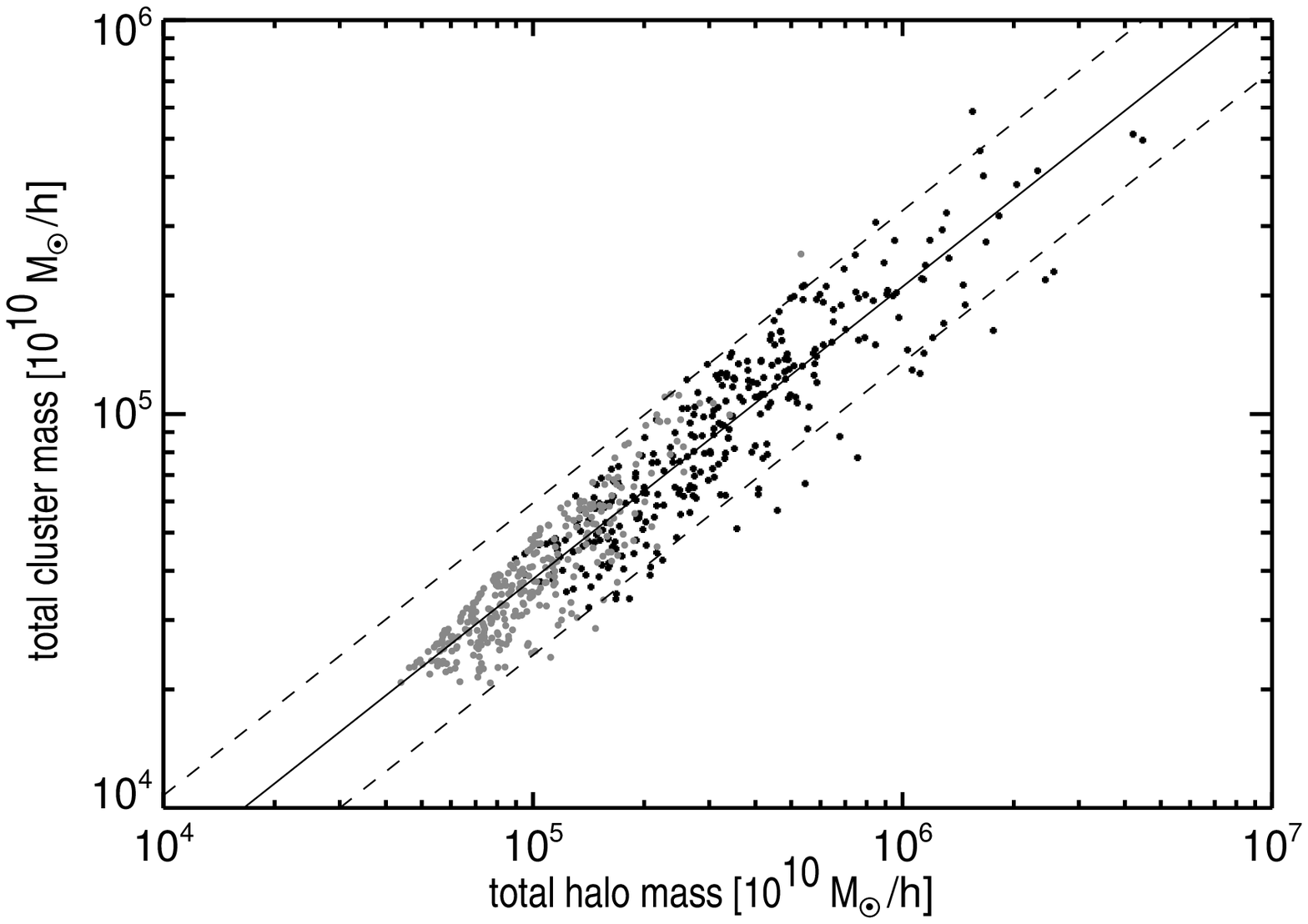}}
  \resizebox{\hsize}{!}{\includegraphics{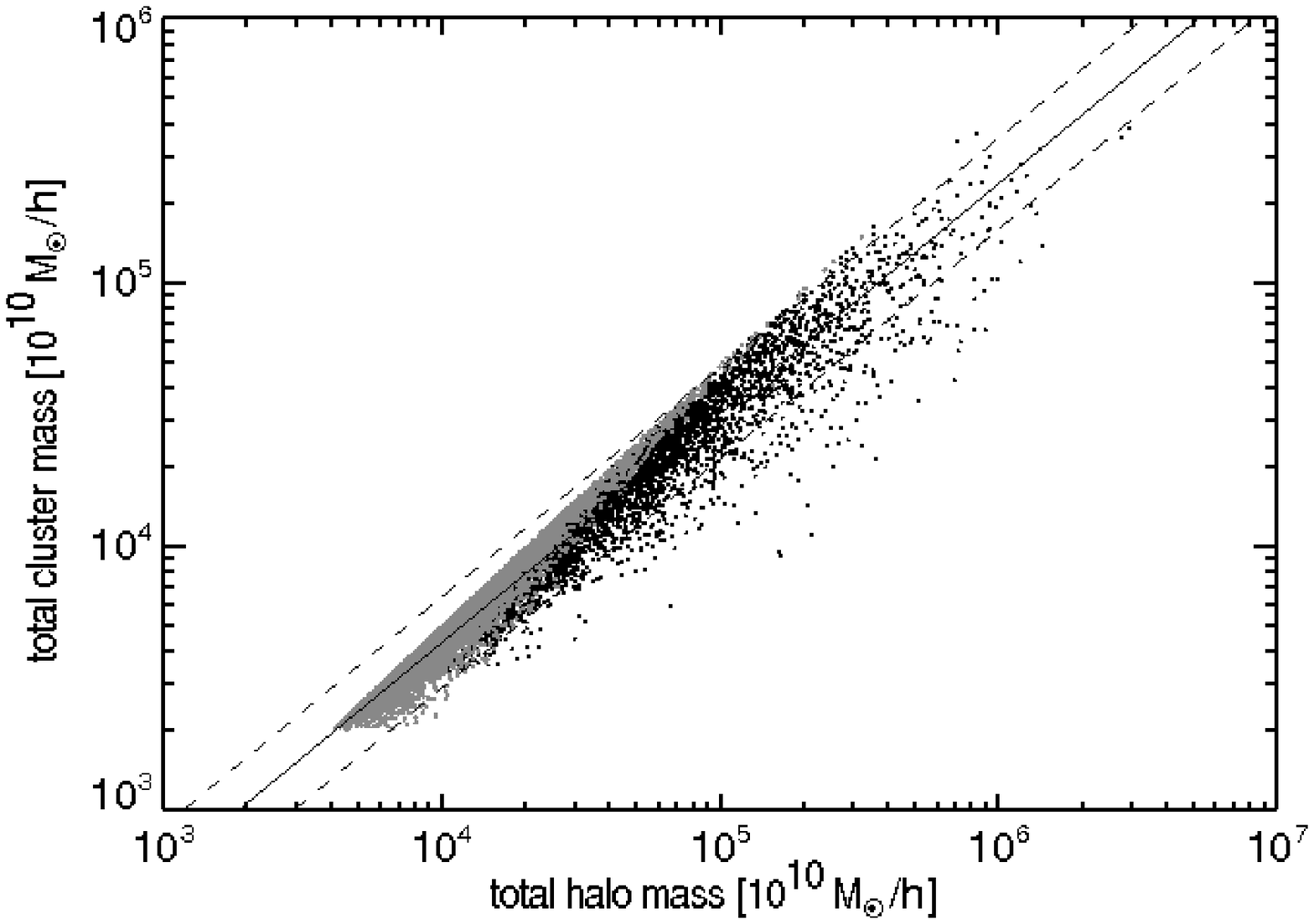}}
  \caption{
    The total mass of a supercluster probed by the total
    mass of member clusters. The sum of the member cluster 
    masses (\textit{CM}) is plotted as a function of the true
    supercluster mass, i.e. the total halo masses (\textit{HM})
    with the correction factor applied (see text for details). 
    The superclusters in the \cs\ catalogue are shown in the
    upper panel, and those in the \gcs\ in the lower one. 
    We show a power-law fit in solid line with 2$\sigma$ scatter 
    of data points in dashed lines.
  }
  \label{fig:mfrac}
\end{figure}

Fig.~\ref{fig:mfrac} shows the comparison of the total mass estimates, CM as a function of 
the HM for the two nominal mass limits, \sfour\ in the upper and \sthree\ in the lower panel. 
In both panels pair superclusters are marked in gray dots and richer superclusters in black dots. 
Note that the lack of scatter at the upper limit in both panels is due to the fact that, 
in particular for close pairs, the cluster masses make up for most of the supercluster
mass and this limit cannot be exceeded. 
This effect is pronounced due to the applied definition of the cluster volume with an extent 
of the supercluster which coincides with its outermost cluster member. 
A more generous volume definition will make this cut-off less sharp. 
This effect also appears in Fig.~\ref{fig:bias}.
A naive expectation is that the mass fraction becomes larger as the mass limit for the 
constituent clusters decreases. 
This is only true if we are considering exactly same superclusters where we can imagine 
lowering the mass threshold to catch more smaller halos. 
However, this does not have to hold if the superclusters vary. 
The mean for this relation for \gcs\ is 0.39 with one sigma scatter of 0.077, and that for 
\cs\ is 0.34 with 0.092. 
As to our expectation the mass fraction decreases with increasing mass limit of the cluster 
catalogue. 
We fit a power-law shown in a solid line in Fig.~\ref{fig:mfrac} just to guide the eye. 
On each side of the solid line two sigma scatter of the data points is shown in dashed lines. 
The power-law model that we used to fit is defined by $Y=B(X/X_o)^A$ where X is the HM, 
Y the CM, $X_o=2\times10^{15}$\smass\ the pivot point, and A and B are the fitted slope and 
amplitude. 
We apply the same weight over all mass range for the Chi-square fit. 
The fitted slope is less than unity over the entire mass ranges, and it decreases towards a 
larger mass limit. 
We note that removal of percolating superclusters does not change this result as they have
a negligible contribution to the total number of systems.
Given the uncertainty in some of the assumptions made above, we refrain here from using the 
results for a determination of supercluster masses, till we have explored the supercluster 
properties in even more detail. 
Our finding is that the total cluster mass is closely correlated with the total halo mass of a supercluster 
with a power-law relation, and that the mean scatter of the relation is less than 40 per cent for all 
cluster mass limits. This result looks very promising for the application of estimating
the masses of superclusters. 
More needs to be investigated especially for alternative volume definitions to truly exploit 
the potential of this approach in the future work.

\subsection{Cluster density bias against Dark Matter}

The rare density enhancements in our Universe are traced by clusters of galaxies. 
Clusters have an advantage compared to other probes of large-scale structure, since fluctuations in their density 
distribution are more biased compared to the DM, which allows us to trace the DM distribution 
very sensitively.
Biasing means in this case, that the amplitude of the density fluctuations in the cluster 
distribution is higher by a roughly constant factor compared to the amplitude of DM fluctuations. 
We studied the bias for the REFLEX II clusters in~\cite{r2ps}, calculating the theoretically 
expected bias based on the formulae given by~\cite{tinker-10}, and tested the results against 
N-body simulations with good agreement for the flux-limited and the Volume-Limited-Sample (VLS) of \rtwo. 
For this study the volume-limited results are relevant. 
For the lower luminosity corresponding to a mass limit of \sfour\ we find a bias factor of 3.3 
and for a mass limit of \sthree\ a bias factor of 2.1.

The power spectrum of clusters of galaxies measures the distribution of clusters as a function 
of a scale, where the amplitude ratio of this power spectrum in comparison to the power spectrum 
of the DM is interpreted as a bias that clusters have. 
Analogous to the power spectrum of clusters of galaxies, we take the number overdensity of clusters 
in superclusters as a measure of bias against the DM overdensity, for which we take again the halo 
mass overdensity as a tracer.
This approach makes use of the observable, the number overdensity of clusters, so it can be 
calibrated against a quantity from the simulation, the DM mass overdensity. 
The cluster number overdensity is defined as
\begin{equation}
  \Delta^N = \left(\rho^N-\rho^N_o \right)/ \rho^N_o
\end{equation}
where $\rho^N$ is the number density of clusters in a supercluster
and $\rho^N_o$ is the mean number density of the clusters in the
simulation. 
Similarly the halo mass overdensity is defined by
\begin{equation}
  \Delta^M = \left(\rho^M-\rho^M_o \right)/ \rho^M_o
\end{equation}
where $\rho^M$ is the mass density of halos in a supercluster
and $\rho^M_o$ is the mean mass density of the halos in the
simulation.

\begin{figure}
  \resizebox{\hsize}{!}{\includegraphics{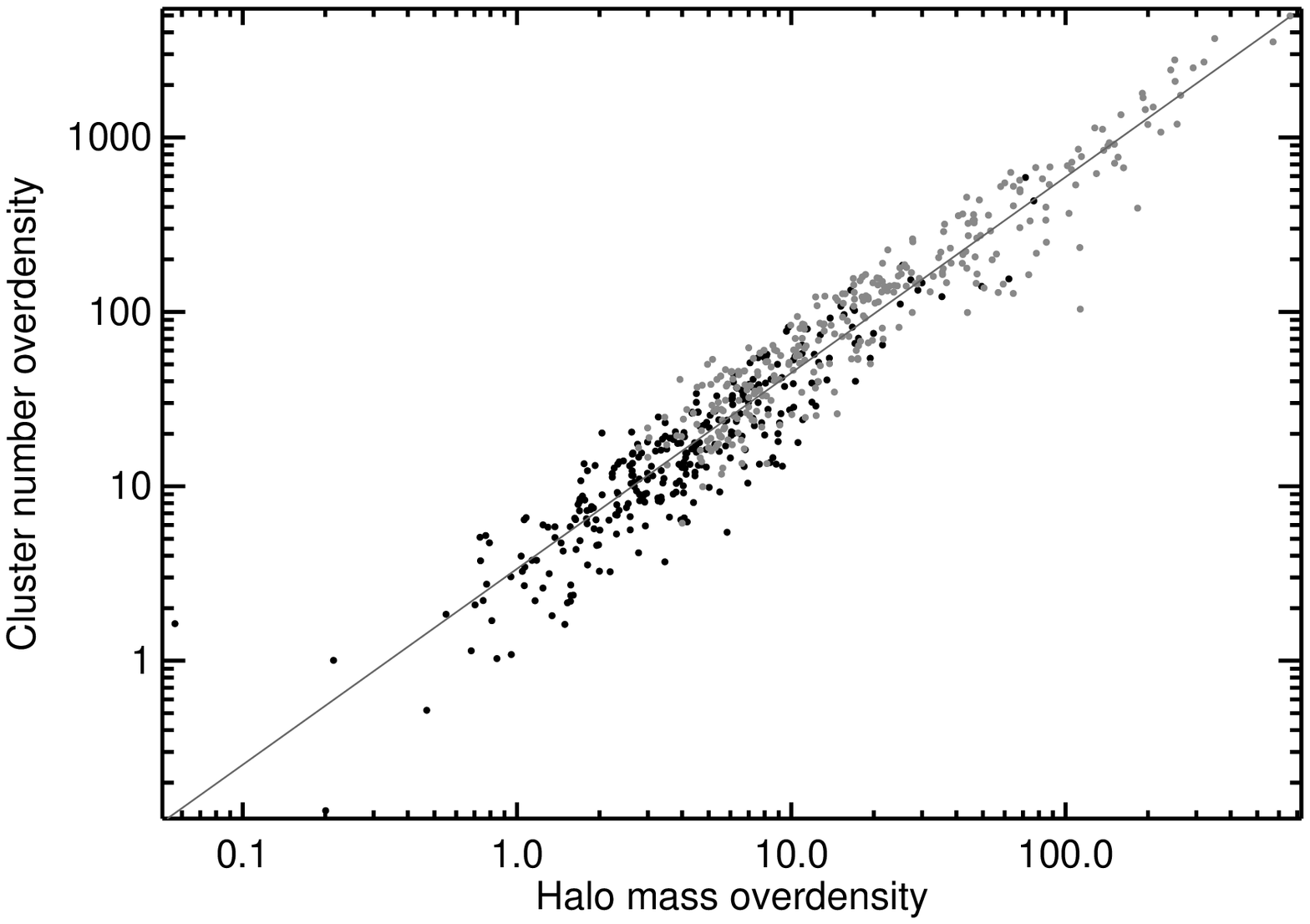}}
  \resizebox{\hsize}{!}{\includegraphics{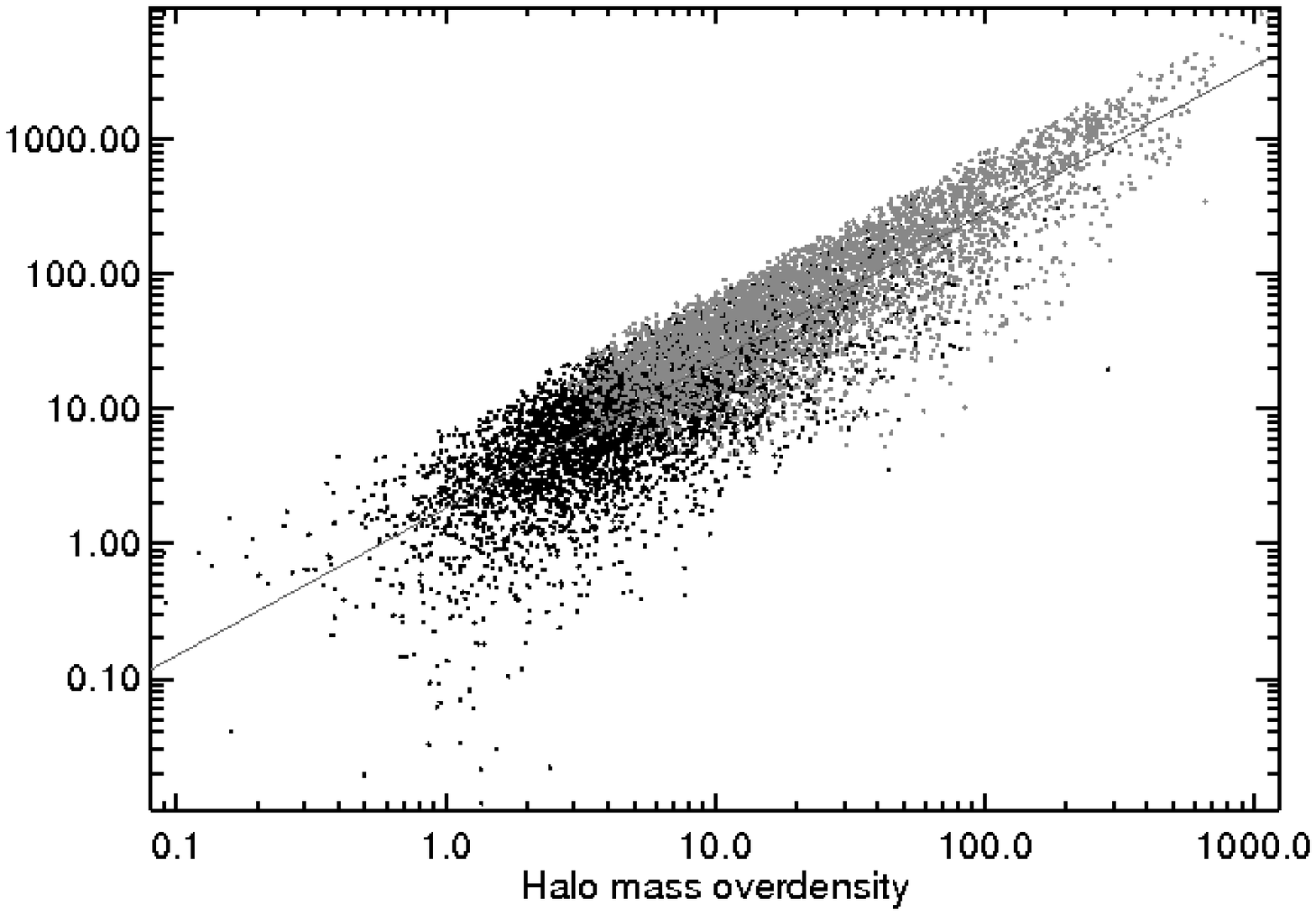}}
  \caption{
    Cluster number overdensity, $\Delta^N$, as a function of the halo
    mass overdensity, $\Delta^M$, for pair (gray dots)
    and richer superclusters (black dots). The best fit model
    (solid line) is shown for the \cs\ (upper) and the \gcs\ 
    (lower panel). The fitted slope and amplitude for 
    different mass limits for the cluster catalogues 
    are shown in Fig.~\ref{fig:bias-amp-00}. 
    }
    \label{fig:bias}
\end{figure}

The cluster number overdensity is plotted as a function of the halo mass overdensity 
in Fig.~\ref{fig:bias} where the upper plot is for the \cs\, and the lower for the \gcs. 
We indicate the best fit of a power-law model with solid lines, which is given by
\begin{equation}
  \Delta^N = B \left( \frac{\Delta^M}{\Delta^M_o} \right)^A
\end{equation}
where $A$ and $B$ are the fitted slope and amplitude, 
$\Delta^M_o$ is a pivot point, which is the median of the 
halo mass overdensity. 

\begin{figure}
  \resizebox{\hsize}{!}{\includegraphics{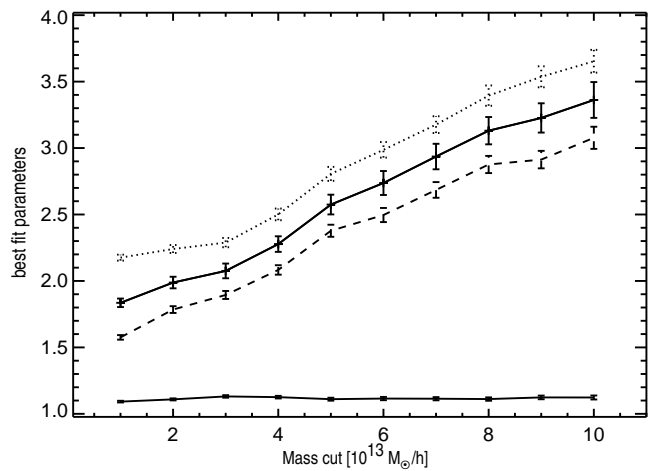}}
  \caption{
    Fitted slope and amplitude for the mass bias as a function
    of the mass limit in the cluster catalogue. 
    The mean of the slope is 1.1, which stays nearly constant over the 
    cluster mass limit range (lower solid line), while the amplitude 
    increases with the increasing mass limit (solid). 
    The errors are calculated with one thousand bootstrapping 
    of the sample. 
    In comparison we show the fitted amplitude of the pair (dotted) and 
    richer (dashed) superclusters separately where the slope is fixed 
    to the fitted slope for the entire sample.  
    }
  \label{fig:bias-amp-00}
\end{figure}

The fitted slope (lowest curve) and amplitude (three upper curves) as a function 
of limiting mass are shown in Fig.~\ref{fig:bias-amp-00}.
The errors are calculated by one thousand bootstrap simulations where we randomly
resample the halo mass overdensity. 
The mean of the slope remains nearly constant at around 1.1 with a very small scatter 
for all mass limits. 
For purely linear bias we would expect a slope of unity. 
That we find a slope of 1.1 shows that any non-linear effects must be very moderate and 
the linear bias picture works as a good approximation. 
Thus we proceed to interpret the scale factor, $B/(\Delta^M_o)^A $, that relates $\Delta^N$ 
and $\Delta^M$ as a bias. 
We find that the bias is 1.83 for the \sthree\ limit, and 3.36 for the \sfour.
This is in reasonable agreement with the expected bias calculated from the cluster
power spectrum for the \rtwo\ quoted at the beginning of this section. 
We note that completely removing the percolating objects identified in Sec. 2.1.3 modifies 
the best fit by 0.3-0.5\% in all mass ranges, hence this result is robust. 

A closer look to the data shows that pair superclusters have a higher effective bias than 
richer systems. 
Due to the fact that pairs contain a large fraction of extremely compact objects we also 
calculated the median bias for the systems richer than pairs. 
Hence we consider the pairs and richer systems separately by fitting the amplitude for 
each population while fixing the slope of the relation to that found above. 
The result of the fit is shown in Fig.~\ref{fig:bias-amp-00} where the amplitude for pairs 
is marked as a dotted line and that for richer superclusters in a dashed line. 
The fair agreement between the cluster biases calculated by the power spectrum and the mass fraction 
approaches is encouraging because the cluster number overdensity clearly extends into the extreme 
non-linear regime with overdensities up to 1000 whereas the calculated bias of the power spectrum is 
mainly based on linear theory. 
This result motivates the study in Sec. 5 where we will test quantitatively how much 
this non-linear environment differs from the field.

\section{Supercluster environment}

To answer the question, how rare superclusters are in the Universe, we study in this section 
the volume fraction that superclusters occupy in the simulation volume. This fraction is 
equivalent to the probability for a randomly chosen point to lie in superclusters.
The rareness and the cluster overdensity in superclusters characterise superclusters as a 
special environment. Thus we study here, if we find different cluster properties in the dense 
supercluster environment compared to the field in terms of a mass distribution, which will 
then be compared to the X-ray luminosity functions obtained with the \rtwo\ superclusters.

\subsection{Volume fraction}

The first quantity that we measure is the volume occupied by superclusters and we calculate 
the volume fraction, which is defined as a ratio of the total volume of superclusters to the 
simulation volume. 

\begin{figure}
  \resizebox{\hsize}{!}{\includegraphics{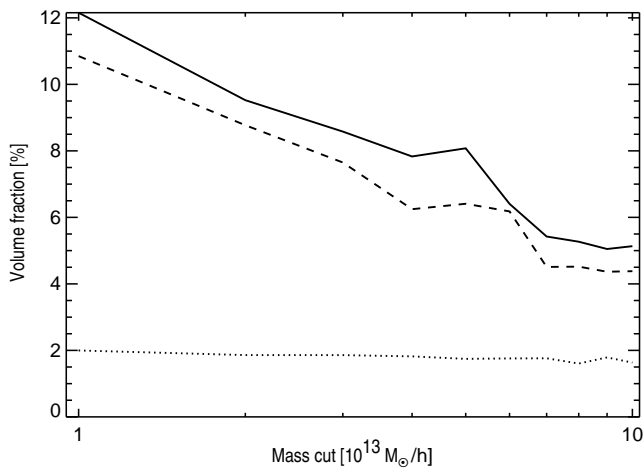}}
  \caption{
    Volume fraction of superclusters in the simulation
    as a function of a limiting mass.
    Richer (solid) and pair (dotted) superclusters are shown. 
    The pair fraction is multiplied by a factor of five for clarity. 
    In comparison we also show the volume fraction of rich 
    superclusters by removing the total volume of possibly percolating superclusters 
    (dashed). 
    The total supercluster volume is completely dominated by 
    the volume of the richer systems, and the pairs occupy 
    an almost constant volume fraction over the mass range. 
  }
  \label{fig:vf}
\end{figure}

Fig.~\ref{fig:vf} shows the volume fraction of rich superclusters in solid and pairs 
in dotted lines, as a function of a limiting mass. 
The volume fraction of pair superclusters is multiplied by five for clarity.
We also explore the contribution from percolated systems to the volume of rich
superclusters, by removing it completely as shown in a dashed line. 

For the lowest mass limit the volume fraction of all superclusters is roughly 12.5 per cent 
while at the \sfour\ limit, the fraction decreases to less than 5.4 per cent. 
In most of the mass range there are typically about five superclusters which may be responsible 
for the percolation, and the range of volume fraction of those superclusters is 3 to 
20~\% of the total supercluster volume. 
This is the largest effect of percolation seen in the properties that are considered in this paper. 
It is not surprising that the majority of the total supercluster volume is made up of the 
volume of richer systems. The volume taken up by the pairs is negligible in comparison, 
and stays nearly constant over the entire mass range. This is due to the fact that the construction 
of pairs only depends on the linking length, which depends on the number density of the clusters 
in the volume, and the distance between two cluster members. 
There is a large fraction of pairs with distances much smaller than the linking length. 
It is interesting to note that the number fraction of clusters, defined as the number of 
clusters in superclusters to the total number of clusters in the simulation, is 
55 per cent for \cs\, and increases to 64 per cent for \gcs.
The number of pair superclusters is roughly half the total number of superclusters. 

This finding is in line with the observational result reported in~\citetalias{chon13}
for the VLS of \rtwo. The volume fraction of superclusters in this sample turned out 
to be two per cent while slightly more than half of the clusters belong to superclusters. 
The X-ray luminosity limit of the VLS is 5$\times10^{43}$ erg~s$^{-1}$ approximately comparable to 
a mass limit of $2\times$\sfour\ below redshift $z=0.1$ for the scaling relation in~\cite{r2cosmo}.
Considering this limiting mass of the VLS, we find that the corresponding volume fraction of 
the simulation is 3.4 per cent.
Both the observation and simulation results show how rare these superclusters are, and 
that they could lend themselves as interesting study objects to probe the non-linear 
regime of the universe.

\subsection{Mass function}

One of the important findings of~\citetalias{chon13} is that the luminosity function 
of superclusters in the VLS compared to that of the field is top-heavy, meaning that 
there are more luminous clusters in superclusters than in the field. 
This implies that superclusters provide a special environment for the non-linear structures 
to grow. 
One possible reason put forward by~\citetalias{chon13} is that X-ray luminosity is tightly 
correlated with the mass of a cluster. This top-heavy luminosity function implies the 
top-heaviness of a mass function of clusters in superclusters. 
An alternative reason could also be an increased frequency of the cluster merger rate due to 
enhanced interactions of clusters where the central region of the cluster is temporarily
compressed giving rise to an increased X-ray luminosity.
To numerically quantify the difference in the X-ray luminosity function, \citetalias{chon13} 
subjected the cumulative X-ray luminosity function to a Kolmogorov-Smirnov (KS) test. Here 
we form equivalently the cumulative mass function with the simulation. 

\begin{figure}
  \resizebox{\hsize}{!}{\includegraphics{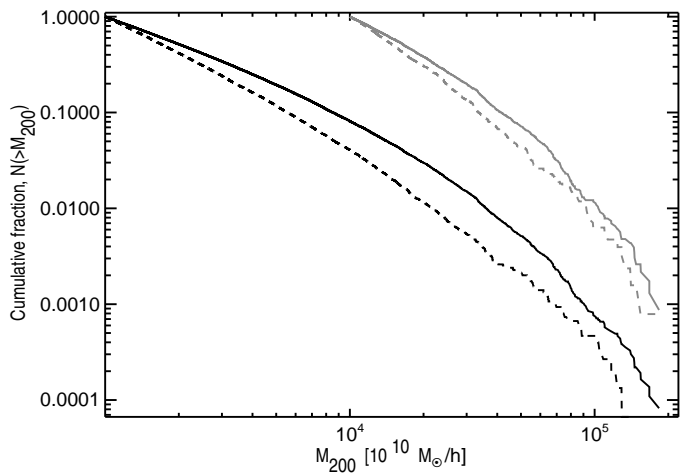}}
  \caption{
    Cumulative mass fraction of clusters inside (solid) and outside 
    the superclusters (dashed) for the \cs\ (gray) and \gcs. 
  }
  \label{fig:lxfunc}
\end{figure}

Fig.~\ref{fig:lxfunc} shows the cumulative mass functions of the two populations of clusters, 
those in superclusters (solid) and those in the field (dashed) for the \cs\ in gray and
\gcs.
To compare these two unbinned cumulative mass functions we use the KS test as in~\citetalias{chon13} 
to calculate the significance of the difference in the two distributions. 
The KS test shows that the probability of both distributions drawn from the same function is 
4.4$\times10^{-10}$ for the \cs\ catalogue of Fig.~\ref{fig:lxfunc}, 
and zero for the \gcs.   
This confirms with very little doubt that the mass function of clusters in superclusters is 
top-heavy in comparison to that of clusters in the field. 
This conclusion is in line with what was observed with the \rtwo\ clusters in superclusters.
The same KS test showed the probability of 0.03 for the \rtwo\ superclusters in the VLS. 
The much smaller KS probabilities resulting from the simulation is due to very large number 
statistics in comparison to the observations. We note that Fig.~\ref{fig:lxfunc} also spans 
about a decade larger in the ordinate than Fig. 10 in~\citetalias{chon13}. 
Hence we confirm that both the X-ray luminosity function and the mass function of clusters 
in superclusters are top-heavy. 
This is an important evidence that the boost seen in the X-ray luminosity of clusters in an enhanced 
density region is not just a temporary phenomenon in merging events, but that the mass function 
of clusters is fundamentally modified so that both mass and luminosity functions in a high density 
environment evolve differently from those of the field.

\section{Discussion and conclusions}

We tested our concept to understand superclusters constructed with a friends-of-friends method
from a complete sample of mass-selected or X-ray luminosity selected galaxy clusters by means
of dark matter halos from cosmological N-body simulations by applying a set-up equivalent
to that used in a typical cluster survey. 
A friends-of-friends algorithm was applied to DM halos above a mass threshold to construct 
a supercluster catalogue. 
One key parameter in this method is the linking length, which is inversely proportional to 
the overdensity parameter, $f$.  
With the simulation we were able to calculate in retrospect the distribution of the number 
density ratio of clusters in superclusters compared to the mean density in the simulation volume. 
As a first-order approach motivated by a theoretical spherical collapse model the geometry 
of a supercluster was assumed to be spherical, and the majority of superclusters turned 
out to have their density ratio greater than the initial input overdensity with 
some scatter due to a particular spatial distribution of clusters within a supercluster.
This suggests that the fof algorithm provides a reasonable way to construct superclusters, 
which approximately selects superclusters with pre-defined cluster overdensity. 

Our findings are encouraging for further studies to use superclusters as potential 
tools to probe the properties of the large-scale structure. We considered the fraction 
of mass that is represented in the cluster content of superclusters in comparison to 
the total mass. There is a clear power-law correlation between those two quantities showing 
that the total member cluster mass in a supercluster is a good first-order estimate
of the total mass of a supercluster with a mean scatter of less than 40 per cent. 
One of our aims in this paper was to learn how the dark matter is traced by superclusters. 
By comparing the cluster overdensity to the underlying DM overdensity, we verified that 
the biasing concept as implied in~\cite{chon13} works. Thus we can indeed 
select superclusters with prescribed DM overdensity with moderate scatter. 
In a similar study we examined the cluster number overdensity as a function of the supercluster 
matter overdensity.
The ratio of these two quantities is the bias carried by the clusters in superclusters. 
We expect that this should closely follow the expectation of the bias for the power spectrum 
of clusters. 
We find that the relation between the cluster number overdensity and the matter overdensity 
as a function of the matter overdensity is described well with a power-law model where the 
slope is within 10~\% of unity and the amplitude increases as expected from theory with 
increasing lower mass limit. 
We interpret the scaling factor between the cluster number density and the halo mass 
density resulting from the power-law model fit as a bias. For superclusters constructed 
with the \cs\ the bias is 3.36 while the bias is 1.83 for the \gcs\ supercluster catalogue, 
which indicates agreement with theory. 
This implies that the similar bias factors which describe global statistical function like the
power spectrum or the correlation function also applies to local overdensities. 
Hence the cluster bias that we found reflects once more that superclusters are a special region 
in our Universe. 

This led to the work in Sec. 5 where we tested how the supercluster environment differs
from the field by considering the volume that superclusters occupy and the mass function of clusters 
in superclusters. 
The volume fraction of superclusters for the VLS-equivalent catalogue is 3.4 per cent in 
good agreement with the result of the \rtwo\ observations in~\citetalias{chon13}.  
It is only a very small fraction that superclusters occupy. 
A high density region in our Universe is expected to evolve differently from the background 
cosmology, and it can be understood as a local universe that evolved from an originally higher 
mass density. 
This can be tested by examining the mass function of clusters in superclusters compared to that 
in the field. 
We find with a close to zero probability that these two mass functions are drawn from the same 
parent distribution, confirming that the mass function is top-heavy analogous to the top-heavy 
X-ray luminosity function in~\citetalias{chon13}. 
In both observation and simulation there are more luminous or more massive clusters in 
superclusters than in the field. 
Hence we confirm with the volume fraction and the mass function of clusters of galaxies 
from the simulation, also supported by the \rtwo\ observations, that the supercluster 
environment is distinctly different from the rest of the universe. 
For the first time this finding is based on the well-understood selection functions of 
clusters in both simulations and large flux-limited X-ray survey data. 
It is also in agreement with previous studies, albeit using different tracer objects to probe 
superclusters and different methods to construct them, such 
as~\citet{einasto03_ref,einasto05_ref,einasto12_ref}. 
The NORAS~II catalogue, complementary to the \rtwo\ survey in the Northern sky, is currently 
being compiled. 
With this addition the sample will effectively double the current sample size, and will 
provide an improved ground for further exploration of supercluster properties.

\section*{Acknowledgments}

We thank the referee for the interesting and constructive comments. 
We acknowledge support from the DfG Transregio Program TR33 and the DFG cluster of excellence 
``Origin and Structure of the Universe'' (www.universe-cluster.de). 
GC acknowledges the support from Deutsches Zentrum f\"ur Luft 
und Raumfahrt (DLR) with the program 50 OR 1305. 
The Millennium Simulation databases used in this paper and the 
web application providing online access to them were constructed 
as part of the activities of the German Astrophysical Virtual 
Observatory. 
We thank Gerard Lemson for his support with the simulation data.

\footnotesize{
  \bibliographystyle{aa}
  \bibliography{scl}

\begin{thebibliography}{38}
\expandafter\ifx\csname natexlab\endcsname\relax\def\natexlab#1{#1}\fi

\bibitem[{{Abell}(1961)}]{abell-61}
{Abell}, G.~O. 1961, \aj, 66, 607

\bibitem[{{Araya-Melo} {et~al.}(2009){Araya-Melo}, {Reisenegger}, {Meza}, {van
  de Weygaert}, {D{\"u}nner}, \& {Quintana}}]{araya09}
{Araya-Melo}, P.~A., {Reisenegger}, A., {Meza}, A., {et~al.} 2009, \mnras, 399,
  97

\bibitem[{{Bahcall}(1988)}]{bahcall-88}
{Bahcall}, N.~A. 1988, \araa, 26, 631

\bibitem[{{Bahcall} \& {Soneira}(1984)}]{bahcall-84}
{Bahcall}, N.~A. \& {Soneira}, R.~M. 1984, \apj, 277, 27

\bibitem[{{Balaguera-Antol{\'{\i}}nez}
  {et~al.}(2012){Balaguera-Antol{\'{\i}}nez}, {S{\'a}nchez}, {B{\"o}hringer},
  \& {Collins}}]{r2ps}
{Balaguera-Antol{\'{\i}}nez}, A., {S{\'a}nchez}, A.~G., {B{\"o}hringer}, H., \&
  {Collins}, C. 2012, \mnras, 425, 2244

\bibitem[{{Batuski} \& {Burns}(1985)}]{batuski-85}
{Batuski}, D.~J. \& {Burns}, J.~O. 1985, \aj, 90, 1413

\bibitem[{{Bogart} \& {Wagoner}(1973)}]{bogart-73}
{Bogart}, R.~S. \& {Wagoner}, R.~V. 1973, \apj, 181, 609

\bibitem[{{B{\"o}hringer} {et~al.}(2014){B{\"o}hringer}, {Chon}, \&
  {Collins}}]{r2cosmo}
{B{\"o}hringer}, H., {Chon}, G., \& {Collins}, C.~A. 2014, ArXiv e-prints

\bibitem[{{B{\"o}hringer} {et~al.}(2013){B{\"o}hringer}, {Chon}, {Collins},
  {Guzzo}, {Nowak}, \& {Bobrovskyi}}]{r2const}
{B{\"o}hringer}, H., {Chon}, G., {Collins}, C.~A., {et~al.} 2013, \aap, 555,
  A30

\bibitem[{{B{\"o}hringer} {et~al.}(2012){B{\"o}hringer}, {Dolag}, \&
  {Chon}}]{scalingpap}
{B{\"o}hringer}, H., {Dolag}, K., \& {Chon}, G. 2012, \aap, 539, A120

\bibitem[{{Chon} \& {B{\"o}hringer}(2012)}]{chon12}
{Chon}, G. \& {B{\"o}hringer}, H. 2012, \aap, 538, A35

\bibitem[{{Chon} {et~al.}(2013){Chon}, {B{\"o}hringer}, \& {Nowak}}]{chon13}
{Chon}, G., {B{\"o}hringer}, H., \& {Nowak}, N. 2013, \mnras, 429, 3272

\bibitem[{{D{\"u}nner} {et~al.}(2006){D{\"u}nner}, {Araya}, {Meza}, \&
  {Reisenegger}}]{duenner06}
{D{\"u}nner}, R., {Araya}, P.~A., {Meza}, A., \& {Reisenegger}, A. 2006,
  \mnras, 366, 803

\bibitem[{{Einasto} {et~al.}(2007{\natexlab{a}}){Einasto}, {Einasto}, {Saar},
  {Tago}, {Liivam{\"a}gi}, {J{\~o}eveer}, {Suhhonenko}, {H{\"u}tsi},
  {Jaaniste}, {Hein{\"a}m{\"a}ki}, {M{\"u}ller}, {Knebe}, \&
  {Tucker}}]{einasto-sim}
{Einasto}, J., {Einasto}, M., {Saar}, E., {et~al.} 2007{\natexlab{a}}, \aap,
  462, 397

\bibitem[{{Einasto} {et~al.}(2007{\natexlab{b}}){Einasto}, {Einasto}, {Tago},
  {Saar}, {H{\"u}tsi}, {J{\~o}eveer}, {Liivam{\"a}gi}, {Suhhonenko},
  {Jaaniste}, {Hein{\"a}m{\"a}ki}, {M{\"u}ller}, {Knebe}, \&
  {Tucker}}]{einasto-07}
{Einasto}, J., {Einasto}, M., {Tago}, E., {et~al.} 2007{\natexlab{b}}, \aap,
  462, 811

\bibitem[{{Einasto} {et~al.}(1994){Einasto}, {Einasto}, {Tago}, {Dalton}, \&
  {Andernach}}]{einasto-94}
{Einasto}, M., {Einasto}, J., {Tago}, E., {Dalton}, G.~B., \& {Andernach}, H.
  1994, \mnras, 269, 301

\bibitem[{{Einasto} {et~al.}(2001){Einasto}, {Einasto}, {Tago}, {M{\"u}ller},
  \& {Andernach}}]{einasto-01}
{Einasto}, M., {Einasto}, J., {Tago}, E., {M{\"u}ller}, V., \& {Andernach}, H.
  2001, \aj, 122, 2222

\bibitem[{{Einasto} {et~al.}(2003){Einasto}, {Jaaniste}, {Einasto},
  {Hein{\"a}m{\"a}ki}, {M{\"u}ller}, \& {Tucker}}]{einasto03_ref}
{Einasto}, M., {Jaaniste}, J., {Einasto}, J., {et~al.} 2003, \aap, 405, 821

\bibitem[{{Einasto} {et~al.}(2012){Einasto}, {Liivam{\"a}gi}, {Tempel}, {Saar},
  {Vennik}, {Nurmi}, {Gramann}, {Einasto}, {Tago}, {Hein{\"a}m{\"a}ki},
  {Ahvensalmi}, \& {Mart{\'{\i}}nez}}]{einasto12_ref}
{Einasto}, M., {Liivam{\"a}gi}, L.~J., {Tempel}, E., {et~al.} 2012, \aap, 542,
  A36

\bibitem[{{Einasto} {et~al.}(2005){Einasto}, {Suhhonenko}, {Hein{\"a}m{\"a}ki},
  {Einasto}, \& {Saar}}]{einasto05_ref}
{Einasto}, M., {Suhhonenko}, I., {Hein{\"a}m{\"a}ki}, P., {Einasto}, J., \&
  {Saar}, E. 2005, \aap, 436, 17

\bibitem[{{Einasto} {et~al.}(1997){Einasto}, {Tago}, {Jaaniste}, {Einasto}, \&
  {Andernach}}]{einasto-97}
{Einasto}, M., {Tago}, E., {Jaaniste}, J., {Einasto}, J., \& {Andernach}, H.
  1997, \aaps, 123, 119

\bibitem[{{Hauser} \& {Peebles}(1973)}]{hauser-73}
{Hauser}, M.~G. \& {Peebles}, P.~J.~E. 1973, \apj, 185, 757

\bibitem[{{Johnston} {et~al.}(2007){Johnston}, {Sheldon}, {Wechsler}, {Rozo},
  {Koester}, {Frieman}, {McKay}, {Evrard}, {Becker}, \& {Annis}}]{johnston07}
{Johnston}, D.~E., {Sheldon}, E.~S., {Wechsler}, R.~H., {et~al.} 2007, ArXiv
  e-prints

\bibitem[{{Kalinkov} \& {Kuneva}(1995)}]{kalinkov-95}
{Kalinkov}, M. \& {Kuneva}, I. 1995, \aaps, 113, 451

\bibitem[{{Liivam{\"a}gi} {et~al.}(2012){Liivam{\"a}gi}, {Tempel}, \&
  {Saar}}]{liiv-12}
{Liivam{\"a}gi}, L.~J., {Tempel}, E., \& {Saar}, E. 2012, \aap, 539, A80

\bibitem[{{Luparello} {et~al.}(2011){Luparello}, {Lares}, {Lambas}, \&
  {Padilla}}]{luparello11_ref}
{Luparello}, H., {Lares}, M., {Lambas}, D.~G., \& {Padilla}, N. 2011, \mnras,
  415, 964

\bibitem[{{Luparello} {et~al.}(2013){Luparello}, {Lares}, {Yaryura}, {Paz},
  {Padilla}, \& {Lambas}}]{luparello13_ref}
{Luparello}, H.~E., {Lares}, M., {Yaryura}, C.~Y., {et~al.} 2013, \mnras, 432,
  1367

\bibitem[{{Peebles}(1974)}]{peebles-74}
{Peebles}, P.~J.~E. 1974, \apss, 31, 403

\bibitem[{{Pratt} {et~al.}(2009){Pratt}, {Croston}, {Arnaud}, \&
  {B{\"o}hringer}}]{pratt09}
{Pratt}, G.~W., {Croston}, J.~H., {Arnaud}, M., \& {B{\"o}hringer}, H. 2009,
  \aap, 498, 361

\bibitem[{{Rood}(1976)}]{rood-76}
{Rood}, H.~J. 1976, \apj, 207, 16

\bibitem[{{Seljak} \& {Warren}(2004)}]{seljak04}
{Seljak}, U. \& {Warren}, M.~S. 2004, \mnras, 355, 129

\bibitem[{{Shandarin} {et~al.}(2004){Shandarin}, {Sheth}, \&
  {Sahni}}]{shandarin04_ref}
{Shandarin}, S.~F., {Sheth}, J.~V., \& {Sahni}, V. 2004, \mnras, 353, 162

\bibitem[{{Sheldon} {et~al.}(2009){Sheldon}, {Johnston}, {Masjedi}, {McKay},
  {Blanton}, {Scranton}, {Wechsler}, {Koester}, {Hansen}, {Frieman}, \&
  {Annis}}]{sheldon09}
{Sheldon}, E.~S., {Johnston}, D.~E., {Masjedi}, M., {et~al.} 2009, \apj, 703,
  2232

\bibitem[{{Springel} {et~al.}(2005){Springel}, {White}, {Jenkins}, {Frenk},
  {Yoshida}, {Gao}, {Navarro}, {Thacker}, {Croton}, {Helly}, {Peacock}, {Cole},
  {Thomas}, {Couchman}, {Evrard}, {Colberg}, \& {Pearce}}]{springel05}
{Springel}, V., {White}, S.~D.~M., {Jenkins}, A., {et~al.} 2005, \nat, 435, 629

\bibitem[{{Thuan}(1980)}]{thuan-80}
{Thuan}, T.~X. 1980, in Physical Cosmology, ed. R.~{Balian}, J.~{Audouze}, \&
  D.~N. {Schramm}, 277--289

\bibitem[{{Tinker} {et~al.}(2010){Tinker}, {Robertson}, {Kravtsov}, {Klypin},
  {Warren}, {Yepes}, \& {Gottl{\"o}ber}}]{tinker-10}
{Tinker}, J.~L., {Robertson}, B.~E., {Kravtsov}, A.~V., {et~al.} 2010, \apj,
  724, 878

\bibitem[{{West}(1989)}]{west-89}
{West}, M.~J. 1989, \apj, 347, 610

\bibitem[{{Zucca} {et~al.}(1993){Zucca}, {Zamorani}, {Scaramella}, \&
  {Vettolani}}]{zucca-93}
{Zucca}, E., {Zamorani}, G., {Scaramella}, R., \& {Vettolani}, G. 1993, \apj,
  407, 470

\end{thebibliography}
}

\label{lastpage}

\end{document}